%% file: 45062corr.tex
\documentclass{aa}
\usepackage{url}
\usepackage{natbib}               
\bibpunct{(}{)}{;}{a}{}{,}
\usepackage{graphicx}             
\usepackage{amssymb}              
\usepackage[usenames,dvipsnames]{color} 
\usepackage[varg]{txfonts}
\usepackage{relsize}
\usepackage{upgreek}

\include{definitions}
\begin{document}

\title{A robust estimation of the twist distribution in magnetic clouds}

\titlerunning{Robust twist estimation in magnetic clouds}
\authorrunning{Lanabere et al.}

\author{Lanabere V.\inst{1}, D\'emoulin P.\inst{2,3} \and Dasso S.\inst{1,4} 
}

\offprints{V. Lanabere}
\institute{
$^{1}$ Universidad de Buenos Aires, Facultad de Ciencias Exactas y Naturales, Departamento de Ciencias de la Atm\'osfera y los Oc\'eanos (DCAO), Laboratorio de Meteorología del esPacio (LAMP), 1428 Buenos Aires, Argentina, \email{vlanabere@at.fcen.uba.ar, sdasso@at.fcen.uba.ar}\\
$^{2}$ LESIA, Observatoire de Paris, Universit\'e PSL, CNRS, Sorbonne Universit\'e, Universit\'e Paris Cité, 5 place Jules Janssen, 92195 Meudon, France, \email{Pascal.Demoulin@obspm.fr}\\
$^{3}$ Laboratoire Cogitamus, 75005 Paris, France\\
$^{4}$ CONICET, Universidad de Buenos Aires, Instituto de Astronom\'\i a y F\'\i sica del Espacio (IAFE), Laboratorio de Meteorología del esPacio (LAMP),CC. 67, Suc. 28, 1428 Buenos Aires, Argentina, \email{sdasso@iafe.uba.ar}
}

   \abstract  
{Magnetic clouds (MCs) are observed \insitu by spacecraft. The rotation of their magnetic field is typically interpreted as the crossing of a twisted magnetic flux tube, or flux rope, which was launched from the solar corona. }
{The detailed magnetic measurements across MCs permit us to infer the flux rope characteristics.  Still, the precise spatial distribution of the magnetic twist is challenging, and thus is debated. 
}
{In order to improve the robustness of the results, we performed a superposed epoch analysis (SEA) of a set of well observed MCs at 1 au.
While previous work was done using the MC central time,
we here used the result of a fitted flux rope model to select the time of the closest approach to the flux rope axis.
This implies a precise separation of the in- and outbound regions to coherently phase the observed signals.  We also searched for and minimised the possible biases such as magnetic asymmetry and a finite impact parameter.   }
{We applied the SEA to derive the median profiles both for the flux rope remaining when crossed by the spacecraft and to recover the one present before erosion. In particular, the median azimuthal B component is nearly a linear function of the radius.  More generally, the results confirm our previous results realised without such a deep analysis. The twist profile is nearly uniform in the flux rope core, with a steep increase at the border of the flux rope and with similar profiles in the in- and outbound regions.  The main difference with our previous study is a larger twist by $\sim 20\%$.
}
{} 
\maketitle


\section{Introduction} 
\label{sect_Introduction}

Large-scale magnetic structures released from the solar corona have been observed over several decades with remote observations. These structures, known as coronal mass ejections (CMEs) are the main drivers for space weather effects at or near Earth. The \insitu observations of CMEs in the interplanetary medium are known as interplanetary coronal mass ejections (ICMEs). 
About one-third of ICMEs have the signatures defining a magnetic cloud (MC): enhanced magnetic field strength with respect to ambient values, a monotonic and high rotation of the magnetic field vector, and a low proton temperature \citep{Burlaga81}. The observed coherent rotation in MCs is interpreted as the passage  through the spacecraft of a large-scale twisted magnetic flux tube or flux rope (FR). 

The twist profile inside the FR is still under discussion, with different authors reaching different conclusions. Some authors \citep[e.g.][and references therein]{Burlaga98,Dasso06,Lepping11,Lanabere20} found that the magnetic twist profile is comparable to the one of the Lundquist model \citep{Lundquist50}.
In this model the twist is nearly constant around the FR centre, increasing from the core outwards, and reaching significant large values towards the FR boundary, and theoretically it can diverge depending on where the end of the FR is defined. Another used model to describe the interplanetary FRs is the Gold-Hoyle model, where the twist distributes uniformly along the radius of the FR \citep[e.g.][]{Farrugia99,Dasso03}.
Magnetic data are also least square fitted by various FR models.    
Since the assumed model defines the distribution of the twist, different authors using different models have found different twist distributions
\citep[e.g.][]{Mulligan99, Hidalgo00, Cid02, Nieves18}.

In a recent study, and without the assumption of a magnetic model for the FR, \cite{Lanabere20} found that the twist in a typical MC is distributed nearly uniformly in the FR core, and it increases towards the FR boundaries; in other words, they found the same expected behaviour for the twist as expected from the Lundquist model.
Meanwhile, other authors support the idea that FRs have a highly twisted core enveloped by a less twisted outer shell \citep[e.g.][]{Wang18}. A more detailed discussion about recent twist studies can be found in \citet{Lanabere20} and \citet[][and references therein]{Florido20}.

Remote observations of CMEs show that their radial extension increases as they propagate away from the low corona. This expansion continues in the interplanetary space as deduced by \insitu observations of the proton velocity \citep[e.g.][]{Demoulin08, Gulisano10, Regnault20}. This expansion creates an ageing effect with an FR growing in size, so with a weaker magnetic field, as the spacecraft crosses it.  This introduces a bias in the data by mixing space and time evolution. \citet{Demoulin20} concluded that the expansion effect in the \insitu measurements is not the main origin of the observed asymmetry in the magnetic field profile of MCs. Thus, the effect of expansion on the deduced twist profile is expected to be weak.

Another important process that MCs experience during their travel though the interplanetary medium is erosion, due to magnetic reconnection with the ambient solar wind \citep[e.g.][]{Dasso06,Dasso07,Ruffenach15}. 
This erosion modifies the balance of magnetic flux present in the inbound region compared to the outbound region, as well as the estimated values of magnetic flux and helicity derived from \insitu measurements 
(the inbound (outbound) region corresponds to the time series when the spacecraft is coming nearer to (moving away from) the FR centre). Moreover, the percentage of eroded magnetic flux is very case dependent and so this needs to be taken into account in MC studies.  In particular, from the analysis of two studied-cases, \citet{Pal21} concluded that the erosion has significant effects on the obtained magnetic flux rope twist, as it is expected mainly due to the lack of its external part, which typically is highly twisted.

In a previous work, \cite{Lanabere20} found the typical twist distribution inside MCs, using a superposed epoch analysis (SEA). The SEA is a classical technique used to enhance the common properties of a set of events while minimising the specificity of individual cases.  An MC has a front and a rear boundary separated by a case-dependent time interval. Thus, in order to superpose the data of many MCs, their boundaries are first set to the same normalised time values within the selected events. Next, an SEA of a scalar quantity is carried out by taking the mean or the median of the cases. This defined an SEA time profile outlining the dominant characteristics of the MC selection \citep[e.g.][]{Masias-Meza16}. This can be generalised to ICMEs and to derive the probability distribution of the studied parameter versus time \citep{Janvier19,Regnault20}.  

When a vector field is considered, such as the magnetic field, before performing the SEA, the data should be rotated to a common frame where the vector components have the same physical meaning for all the studied cases. Since MCs have an FR structure, the FR frame is the relevant frame, with the condition that all FRs need to be set to the same helicity sign so that the magnetic signals add up. Setting the same impact parameter sign is also needed.
Then, an SEA of the axial and azimuthal components can be performed, from which a twist profile is derived \citep{Lanabere20}.  These authors applied this technique to a subset of MCs characterised by low asymmetry in the $B$ profile to minimise its effect. This resulted in a typical twist profile that was nearly uniform in the FR core, and which moderately increased towards the MC boundaries.

However, by using the MC boundaries for the SEA, \citet{Lanabere20} did not consider the in- and outbound asymmetries typically present in MCs (e.g. those produced by the erosion).
Thus, such an SEA mixes MCs that have a different imbalance of magnetic flux in the in- and outbound sides.  Said differently, the normalised time corresponding to the closest approach distance of the spacecraft to the FR axis would be at a different location for each of the superposed MCs. This could especially affect the deduced azimuthal component profile, which is reversing at the FR axis, and then this could impact the deduced twist profile. 

\cite{Demoulin19} presented a new analysis of the results obtained by \cite{Lepping90}, with consequences on the interpretation of the FR boundaries. 
A reference time is defined for the minimum approach distance of the spacecraft to the fitted FR axis, which in general is different to the central MC time. In general, this implies a different size of the FR in the in- and outbound regions. 
Then, the same size can be selected on both sides in order to get an FR. This allows us to superpose MC data limited to an FR extension with the closest approach time of FR axis for all cases located at the same time. Finally, this enhances the physical coherence of the superposed field components of the MC data. 

The above improvements of the SEA of MC data is done in steps, as follows. In \sect{Data} we describe the data and the studied quantities. We define a method to assign a sign to the asymmetry factor reported in Lepping's catalogue so that the fitted FR is fully defined.  In particular, this defined the central reference time corresponding to the FR axis for each studied MC. In \sect{method} we implement the procedure developed by \citet{Demoulin19} to define the radius of the FR remaining when crossed by the spacecraft and the radius of the non-eroded FR. Then, we present the SEA applied to different MCs subsets in order to analyse, then to minimise, the identified biases present within the magnetic field components. Next, in \sect{SEA_twist} we present the SEA twist profiles of the eroded and non-eroded FRs. Finally, in \sect{Conclusions} we present a summary and our conclusions.

\section{Data and fitted flux rope model}
\label{sect_Data}

\subsection{Data}
\label{sect_Data_MC}

In this work we used the \cite{Lepping06} catalogue\footnote{\url{https://wind.gsfc.nasa.gov/mfi/mag_cloud_S1.html}}, which
consists of a full description of more than 160 MCs observed by the Wind spacecraft between 1995 and 2012. This catalogue includes physical quantities that were estimated with the Lundquist's FR model fitted to the MC data \citep{Lepping90}.
In particular, we used the start and end time of the passage of Wind through the MCs, the FR orientation $\tL$ $\pL$, the closest approach distance $CA$ of Wind to the FR axis, the asymmetry factor $\asf$, the FR radius $\Ro$ (in $\au$), and the magnetic helicity sign $H$. In this catalogue, each MC is classified into three different categories according to the fitting quality ($\Qo$), where $\Qo = 1$ means that a good fitting was obtained, $\Qo = 2$ stands for fair quality, and $\Qo = 3$ for poor quality, as defined in Appendix A of \citet{Lepping06}. We kept a set of 91 MCs of quality $\Qall$.

Wind spacecraft data were used in this study. In particular, we used data from the Magnetic Field Instrument (MFI)\footnote{\url{https://cdaweb.sci.gsfc.nasa.gov/pub/data/wind/mfi/mfi_h0}} and the Solar Wind Experiment (SWE)\footnote{\url{https://cdaweb.sci.gsfc.nasa.gov/pub/data/wind/swe/swe_h1/}}, with a temporal cadence of 60~s for the MFI and 92~s for the SWE. From these data we used the magnetic field magnitude and components in the GSE coordinate system, and a proton wind speed component along the Sun-Earth direction. 

The study of geometric and physical properties of MCs, such as asymmetry, magnetic flux, and twist, is best defined in the local frame $(\ux ,\uy, \uz)$, known as the orthonormal FR frame. The $Z$ axis is defined by the FR axis direction, oriented by the magnetic field direction in the central FR part. $X$ and $Y$ axes are orthogonal to the $Z$ axis with the spacecraft trajectory contained in the $X,Z$ plane. In the FR frame, the magnetic field components are noted as ($\Bx,\,\By,\,\Bz$) and ($\bx,\,\by,\,\bz$) when normalised by the magnetic field strength.
The observed magnetic field components are rotated from GSE to the FR frame using Lepping's FR axis orientation ($\tL$, $\pL$).  A study of the differences between Lepping's orientations and minimum variance (MV) orientations can be found in \citet{Lanabere20}. In that work, we showed that there are small differences between MV and Lepping orientations for the set $\Qall$ (Pearson and Spearman correlation coefficients about 0.93 and 0.78 for $\theta$ and $\phi$, respectively), leading to comparable SEA profiles.

We quantify the asymmetry of the $B(t)$ profile as in \citet{Lanabere20} with the barycentre of the magnetic field intensity:
    \begin{equation}
    \CB = \frac{\int_{0}^{\Delta t} (t/\Delta t -1/2) \, B(t) \, \rmd t}
               {\int_{0}^{\Delta t}  B(t) \, \rmd t}
     \label{eq_CB},
    \end{equation}
where the observed MC is set in the time interval $[0, \Delta t ]$.
The factor $(t/\Delta t -1/2)$ defines the time fraction to the MC central time. $\CB = 0$ for any  symmetric $B(t)$ profile around the central time, while $\CB <0$ ($\CB >0$) when the magnetic field is stronger on the inbound (outbound) side. Finally, the magnitude of $|\CB |$ measures the importance of the asymmetry.

\begin{figure}[t!]            
\centering
\includegraphics[width=0.49\textwidth, clip=]{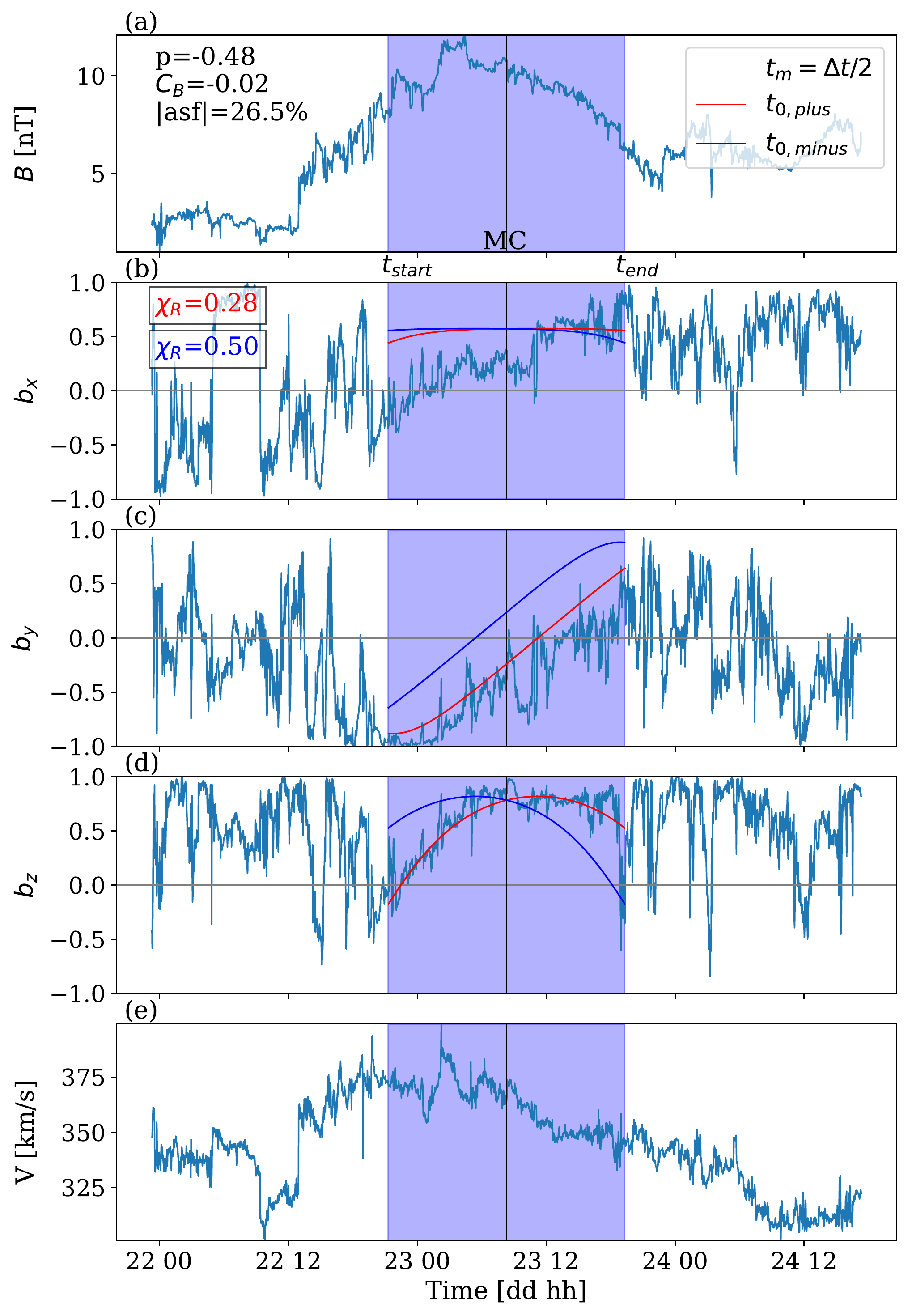}
\caption{Selection of the correct sign of the $\asf$ for the MC observed on 22 August 1995. The MC region is shown with a blue background. (a) The magnetic field magnitude, (b,c,d) the magnetic field components, normalised to the field magnitude, in the FR frame. Two Lundquist solutions fitted to the observed magnetic field components, with their origin set at $\tm$ (blue) and $\tp$ (red) is added in (b,c,d). (e) The observed velocity profile. The correct sign of the $\asf$ is determined by the best fit represented by the lower $\chiR$ value \eqp{chiR}.  Here it is for $\sgn(\asf)=1$ (red curves).}
 \label{fig_event}
\end{figure}

\subsection{Definition of the $\asf$ sign}  
\label{sect_asf_sign}

The fit of the Lundquist's model to the magnetic data provides the time when the spacecraft was the closest from the axis of the best-fitted FR. The asymmetry factor ($\asf$) reported in Lepping's catalogue provides information about how far the centre of the observed MC time interval is with respect to the closest approach time. However, only the absolute value of the $\asf$ is provided.  In this section we define a procedure to retrieve the sign of the $\asf$.  
\figsss{event} shows the procedure applied to an example, the MC observed on 22 August 1995.

The absolute value of the asymmetry factor ($|\asf|$) provided in Lepping's catalogue \citep{Lepping06} allows us to compute the time of the closest approach as
    \begin{equation}
    t_0 = \Delta t/2(1 \pm \rm |\asf|/100) \,.
    \label{eq_to}
    \end{equation}
We added the absolute value to the $\asf$ to reinforce that Lepping provides only positive values. Then, \eq{to} results in $t_0 = \Delta t/2$ when the closest approach of the FR axis is at the centre of the time interval ($\asf = 0\%$), and  $t_0 \neq \Delta t/2$ otherwise. We note the two possible values for $t_0$ in \eq{to} with the minus and plus signs as $\tm$ and $\tp$, respectively. These two options are shown in \fig{event} as blue and red vertical lines for $\tm$ and $\tp$, respectively.

The linear force-free field model with cylindrical symmetry \citep{Lundquist50} is used to fit the observed MC data (blue background in \fig{event}). In cylindrical coordinates ($\ur, \utheta, \uz$), this model writes
    \begin{equation}
    \vBL(R) = \Bo \, J_1(\alpha R)\,\utheta + \Bo \, J_0(\alpha R)\, \uz \,
    \label{eq_BL},
    \end{equation}
where $J_0$ and $J_1$ are the ordinary Bessel functions of order zero and order one, respectively, while $\Bo$ and $\alpha$ are the two free parameters of the model. 

In order to fit the Lundquist's model to the observed magnetic field components in the FR frame, the time $t$ needs to be converted into the spatial coordinate $X$ with its origin set at where the minimum approach distance occurs ($X(t_o)=0$, $X<0$ for the inbound branch, and $X>0$ for the outbound one),
    \begin{equation}
    X(t) = (t-t_0)\,\Vmean\, \rm cos\,\lambda \,,
    \label{eq_X}
    \end{equation}
where $\Vmean$ is the mean solar wind speed of the MC and $\lambda=\sin^{-1}\,(\cos \pL\,\cos \tL)$ is the location angle measured from the plane ($\uyGSE$, $\uzGSE$) towards the MC axis, as defined by \citet{Janvier13}. 
Next, the signed distance to the FR axis, along the spacecraft trajectory inside the FR, is computed as
    \begin{equation}
    R = \sgn(X)\sqrt{X^2 + (p\Ro)^2} \,,
    \label{eq_R}
    \end{equation}
where $p$ is the impact parameter defined as $p=CA/100$ and $\Ro$ is the FR radius from Lepping's table defined by the first zero of $J_0 (\alpha R)$ (i.e. $\Ro$ such that $J_0 (\alpha \Ro)=0$). 

For each option of the time of the closest approach (i.e. $\tm$ or $\tp$), we computed the normalised Lundquist solution
    \begin{equation}
    \vbL(R) = \frac{J_1(\alpha R)}{\sqrt{J_1^2(\alpha R)+ J_0^2(\alpha R)}}\, \utheta 
            + \frac{J_0(\alpha R)}{\sqrt{J_1^2(\alpha R)+ J_0^2(\alpha R)}}\, \uz \,.
    \label{eq_bL}
    \end{equation}
Next, the cylindrical magnetic field components are transformed to the orthonormal FR frame ($\ux,\uy,\uz$).  We recall that $\ux$ is in the plane defined by the spacecraft trajectory and the FR axis ($\uz$).  The Lundquist model writes
    \begin{equation}
    \vbL(R) = \bthetaL \frac{p\, \Ro}{R}\, \ux + \bthetaL \frac{X}{R}\,\uy + \bzl\, \uz \,,
    \label{eq_bL_cartesian}
    \end{equation}
where $X$ is given by \eq{X}, $p\Ro$ is the distance of the spacecraft from the flux rope axis at the closest approach point, and 
$\bthetaL$ and $\bzl$ are the normalised azimuthal and axial magnetic field components from Lundquist solution \eqp{bL}. 

\begin{figure}   
    \centering
\noindent\includegraphics[width=0.49\textwidth]{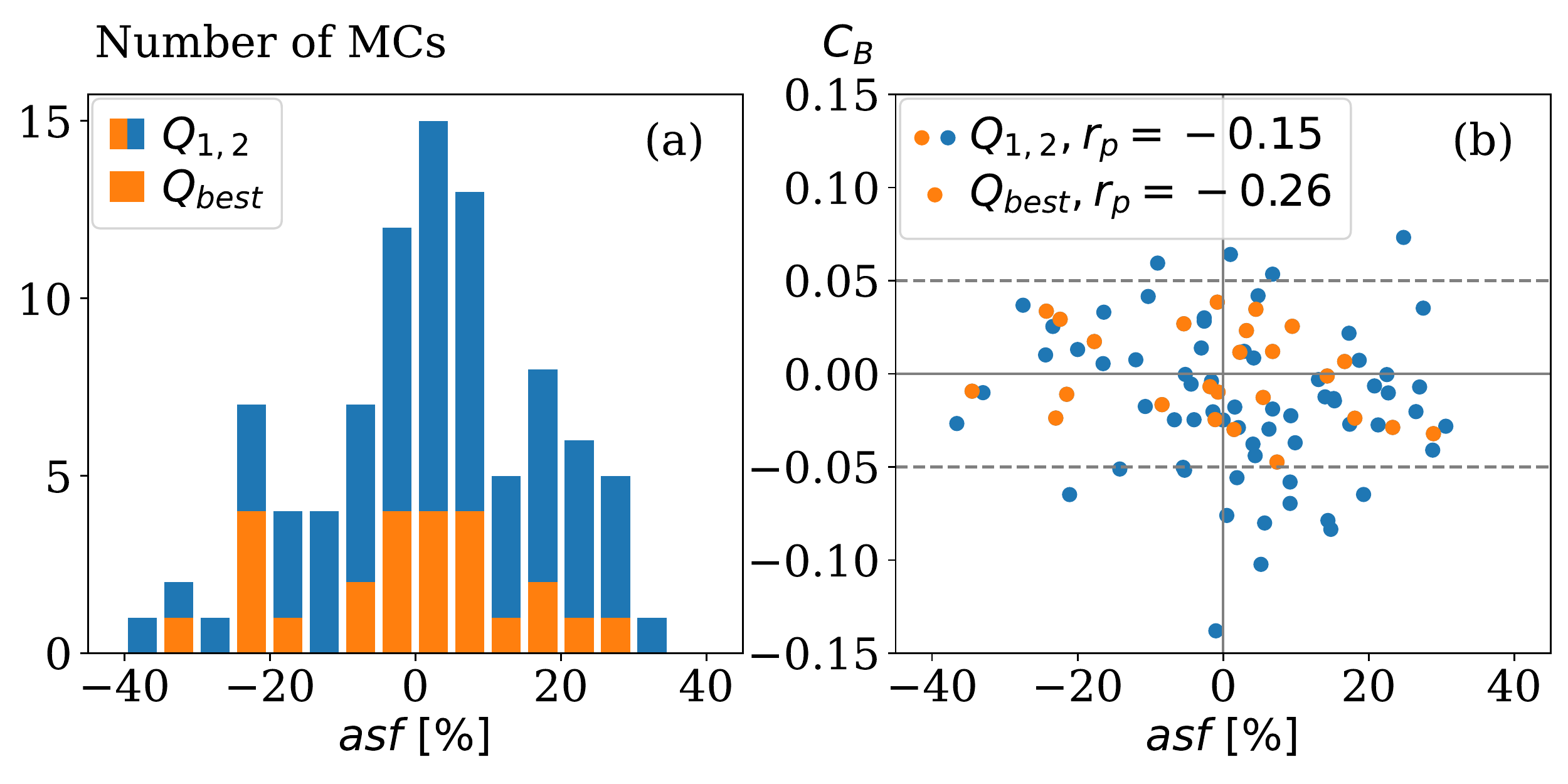} 
    \caption{
    Signed asymmetry factor results. (a) Histogram of the signed $\asf$ (related to the time of the closest approach to the FR axis with \eq{to}).
    (b) Scatter plot of $\asf$ vs $\CB$ (related to the magnetic field profile symmetry). The group $\Qall$ of analysed MCs and its subset $\Qbest$ are shown. The dashed line in (b) corresponds to the most symmetric MCs with $|\CB|\leq 0.05$.}
    \label{fig_asf}
\end{figure}

Next, we computed the two possible models for each MC as the example shown in \fig{event} with blue and red lines corresponding to $\tm$ and $\tp$, respectively.
The quality of the fit between the normalised Lundquist magnetic field components and the normalised observations is characterised by the reduced chi-square defined as
   \begin{equation}
   \chiR = \sqrt{\frac{\left(\bx - \bxl \right )^2 
                       +\left(\by - \byl \right )^2 
                       +\left (\bz - \bzl \right)^2}{3N_d-n}}  \,,
   \label{eq_chiR}
\end{equation}
where $n = 5$ is the number of parameters of the fit and $N_d$ is the number of data points as defined by \citet{Lepping06}. Then, we computed $\chiR$ for both fits (with $\tm$ and $\tp$).

Finally, we designated a sign to $\asf$ according to the fit that presents the lowest $\chiR$. This implies that $\asf > 0$ when the best fit is obtained with $\tp$, and $\asf <0$ when the best fit is obtained with $\tm$. 
As a summary, we present in \fig{asf}a the distribution of the signed asymmetry factor for $\Qall$ and the subset $\Qbest$ defined with $|\CB|\leq 0.05$, and with $|p|\leq 0.3$ \citep{Lanabere20}. We remind the reader that all $\CB$ values reported by \citet{Lanabere20} are larger by a factor of two. In particular, we found 53 cases (58 \% of cases, see Table \ref{sect_Table}) having an $\asf>0$, which corresponds to have more erosion in the MC back, assuming the observed asymmetry is mainly due to erosion. The other 38 cases (42 \% of cases) with an $\asf<0$ correspond to more erosion in the front.

The distribution of $\Qall$ presents a maximum near zero, with a low bias to positive values ($58\%$ positive). Furthermore, there were no cases with $|\asf|>40\%$. Meanwhile, the distribution for $\Qbest$ is more evenly distributed ($52\%$ positive).

The shift of time for the closest axis approach, $t_0$, from the central MC time could be due to the FR erosion during the travel from the Sun. For example, if reconnection at the FR front had removed magnetic flux, $t_0$ is earlier than $\Delta t/2$, or  the $\asf <0$, and also $\CB <0$ as $B(t)$ is stronger in front. On the other hand, erosion at the FR rear implies
$t_0$ later than $\Delta t/2$, the $\asf > 0$, and $\CB >0$.  Another possibility is the asymmetric compression of the FR.  For example, if the inbound region is more compressed (by a strong sheath), $B(t)$ is stronger there, $\CB <0$, and the inbound side is less extended implying the $\asf <0$. 
Thus, if erosion or asymmetric compression of the FR is the main origin of a finite $|\asf|$, we expect a positive correlation  between the $\asf$ and $\CB$.
In fact, \fig{asf}b shows only a weak anti-correlation with the Pearson correlation coefficients slightly negative ( $-0.15$ for $\Qall$ and $-0.26$ for $\Qbest$). 
Therefore, the asymmetry in the magnetic field profile $B(t)$, either by erosion or compression, is not dominantly at the origin of the shift in time of the position of the closest axis approach.

 \begin{figure}    
    \centering
\noindent\includegraphics[width=0.49\textwidth]{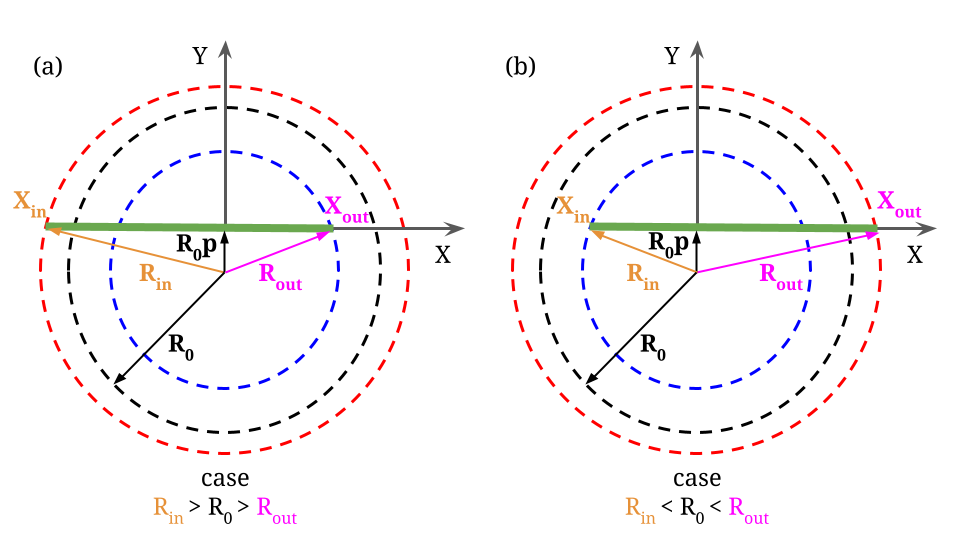} 
    \caption{Schema of an MC cross-section used to define the MC parameters defined by \cite{Demoulin19}. The MC interval, deduced from \insitu observations, is schematised with a thick green line. The dashed circles represent the boundary for three characteristic FRs: in blue for the \insitu FR, in black for the fitted Lundquist FR limited within $\Bz(\Ro)=0$, and in red for the expected fully formed FR without erosion at its boundary. 
    Two cases are shown: (a) $\Rin>\Ro>\Rout$ and (b) $\Rin<\Ro<\Rout$ corresponding to the MC examples shown in \fig{t2R}.
    }
    \label{fig_schema}
\end{figure}
\begin{figure*}[t!]            
\centering
\includegraphics[width=0.49\textwidth, clip=]{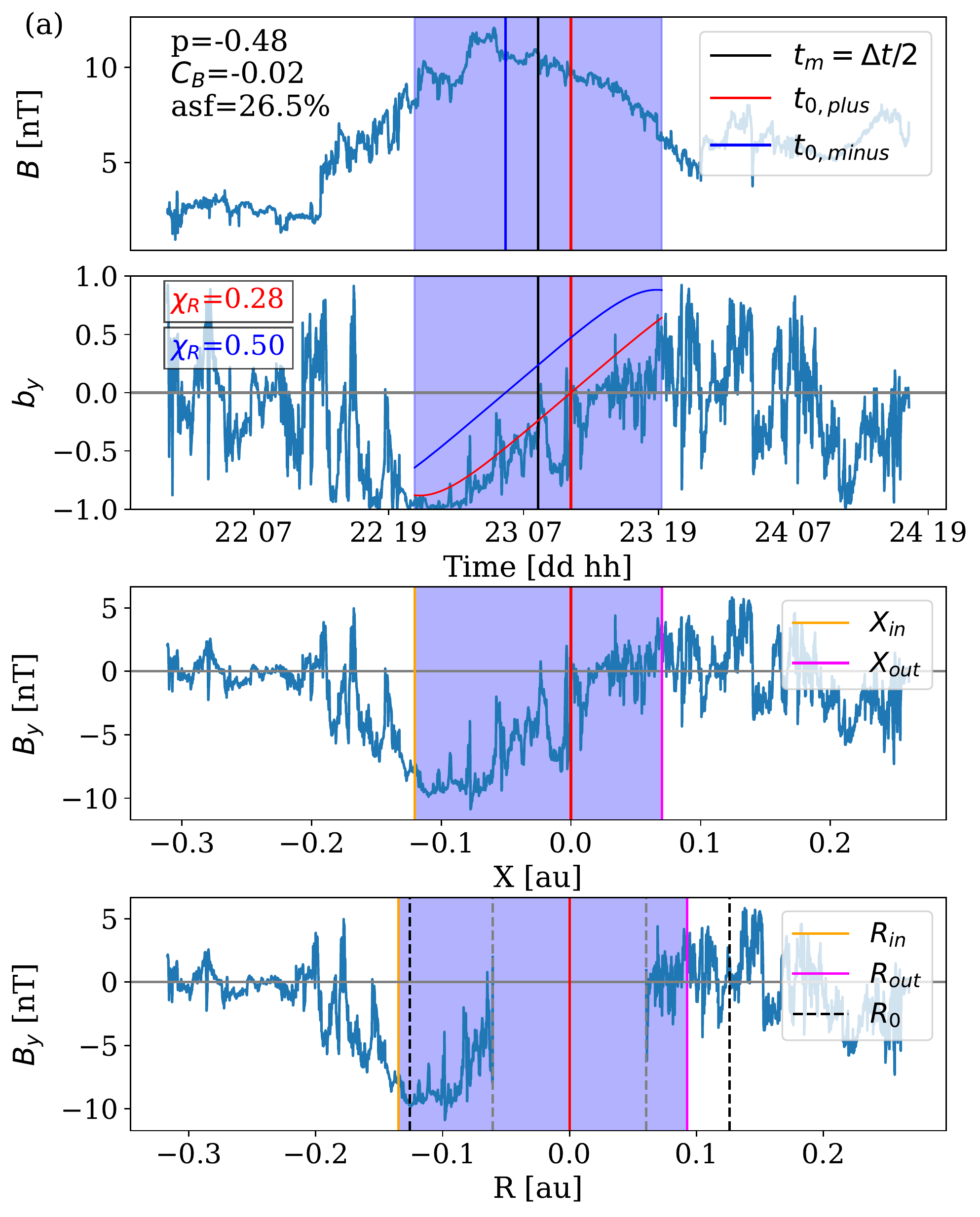}
\includegraphics[width=0.49\textwidth, clip=]{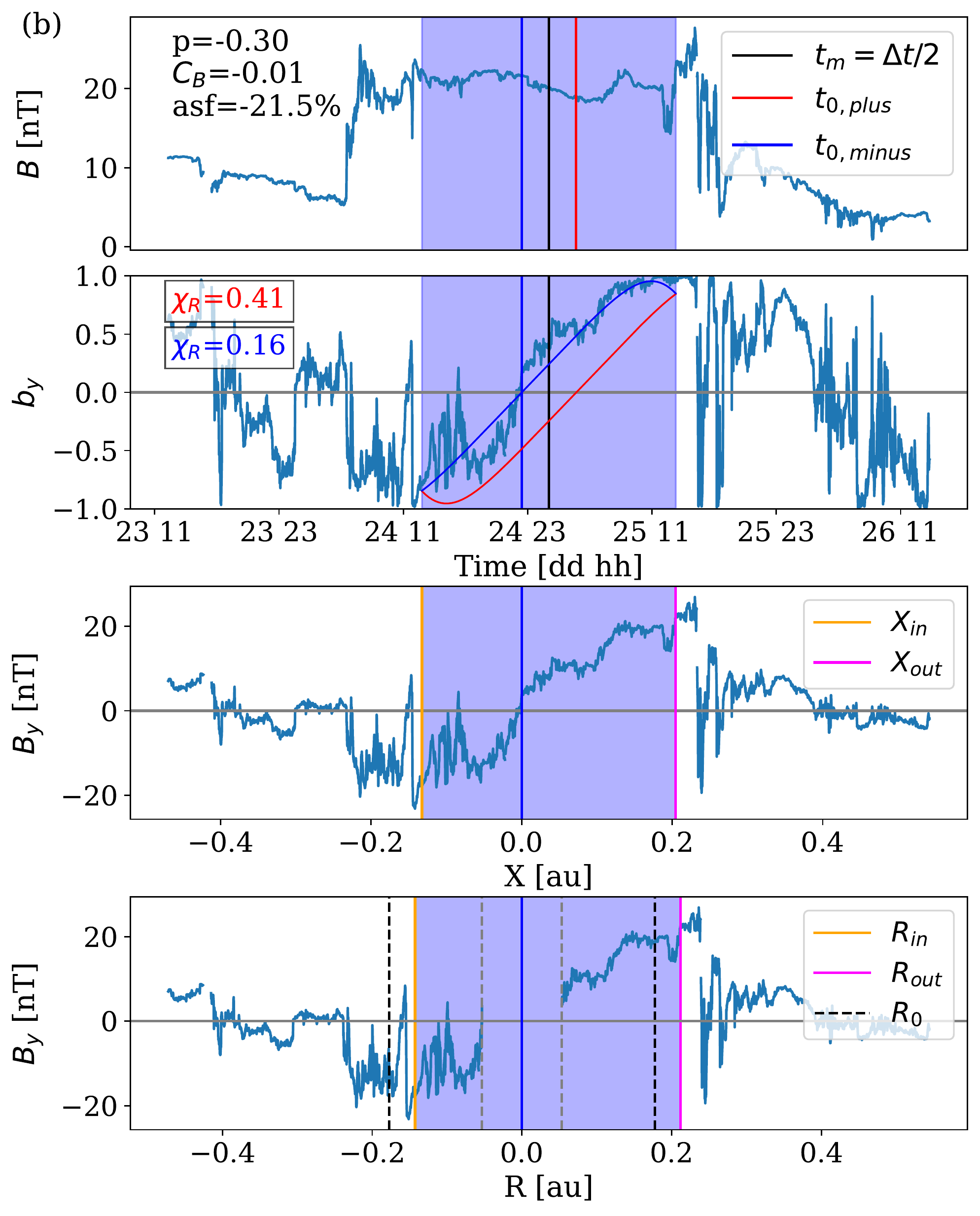}
\caption{Definition of the $\asf$ sign and the in- and outbound regions of the MC in $X$ and $R$ spacial coordinates. From top to bottom, the first two rows show the magnetic field magnitude $B$ and the $\By$ component in function of time. The 
$\By$ profile is dominating in $\chiR$ \eqp{chiR} to determine the sign of the $\asf$.  The third and fourth rows show the in- and outbound regions of the MC in $X$ and $R$ spacial coordinates, respectively.
 Column (a) corresponds to an MC observed on 22 August 1995 as an example of an $\asf>0$ with $\Rin>\Ro>\Rout$ as the schema shown in \fig{schema}a. Furthermore, in this case $\Rinf = \Rout$ and $\Rsup = \Rin$. Column (b) corresponds to an MC observed on 24 July 2004 as an example of an $\asf<0$ with $\Rin<\Ro<\Rout$ as the schema shown in \fig{schema}b.  In this case, $\Rinf = \Rin$ and $\Rsup =\Rout$.}
 \label{fig_t2R}
\end{figure*}

\section{Method} 
\label{sect_method}
 
The main aim of the present section is to improve the SEA of MCs as outlined in \sect{Introduction}. In order that the magnetic field components of the different MCs add up constructively, they need to represent the same quantity (e.g. the axial component) and have the same polarity (\sect{SEA_method}). Moreover, since a spacecraft provides data only on an eroded, or partially formed, FR, we also need to infer a coherent FR for each MC so that the SEA adds up coherently the same FR portion of the data. This FR definition could be realised with the fitted FR model, or with the FR remaining when the MC is observed, or even with the expected fully formed FR (counting on the statistical approach to fill the data gap of individual MCs for the eroded part).  This procedure of defining an FR associated with the data is the object of \sect{FR_radius}. The steps needed to perform an SEA of the vector magnetic field are described in \sect{SEA_method}. Since we identified biases in this SEA, we define MC subsets to decrease them, at the expense of lower statistics (\sect{Data_MC_selection}). Finally, we analyse the various SEA results (\sect{Comparing_SEA_profiles}).

\subsection{Definition of FR radius}
\label{sect_FR_radius}

In order to define an FR from MC data, we applied the procedure described by \cite{Demoulin19}. We first defined the inferior and superior values of the FR radius ($R$) at the MC boundaries, for each MC of the set $\Qall$. This procedure is shown applied to two examples of MCs in \figs{schema}{t2R}.
The MC observed on 22 August 1995 is represented by the schema shown in \fig{schema}a with the time of the closest approach $t_0=\tp$ (see \fig{event} and \sect{asf_sign}). The MC observed on 24 July 2004 with $t_0=\tm$ is represented by \fig{schema}b. 

The observed MCs are typically not compatible with a full FR since the time of the FR axis closest distance, $t_0$, is typically not at the centre of the observed MC time interval. This is linked to the asymmetry of the magnetic field profile due to asymmetric compression and erosion \citep{Dasso07,Ruffenach15}.
More precisely, the amount of azimuthal magnetic flux is not balanced between the in- and outbound regions.  To precisely correct this, the FR boundary could be computed by imposing this flux balance from the data transformed to the FR frame.  
An alternative is to use the fitted FR model to define this flux balance. It has the advantage of filtering out magnetic fluctuations such as the ones created by Alfvén waves.  In any case, the two procedures are expected to provide very close results since we select the set of MCs that are best fitted by the model. The remaining small differences are further minimised by the SEA procedure.

The typical MC asymmetry implies that the fitted FR extends to a radius $\Rin$ and $\Rout$, which are generally different at the in- and outbound MC boundaries. Thus, the model is not an FR ending at a given radius $R$. This could occur when either the MC is formed of a smaller FR, of radius $\Rinf = \min (\Rin, \Rout)$, with extra $B$ flux on one side, or the initial FR was eroded on one side during its travel, so that the total FR would have been $\Rsup = \max (\Rin, \Rout)$ without erosion. These two interpretations are shown in \fig{schema} for the two cases $\Rin > \Rout$ and $\Rin<\Rout$. We precise the derivation of these radius below, with the computation steps summarised in \fig{t2R} for two MCs with  a similar asymmetry factor, impact parameter, and a nearly symmetric $B$ profile.

After the definition of the $\asf$ sign, we computed the spatial coordinate $X$  using \eq{X} with its origin set at $\tp$ or $\tm$ (third row of \fig{t2R}). In this spatial coordinate the in- and outbound boundaries of the MC are located at $\Xin$ and $\Xout$, as defined by \cite{Demoulin19}.
For the MC of \fig{t2R}a the maximum $\chiR$ selects $\tp$, and then the inbound boundary is more extended than the outbound boundary. Meanwhile in \fig{t2R}b, $t_0=\tm$ and the outbound boundary is larger than the inbound one. The MC boundaries are associated with the FR radius $R$ defined as \eqp{R}:
   \begin{equation}
   \Rin  = \sqrt{\Xin^2 + (p\Ro)^2} \quad \rm{and} \quad 
   \Rout = \sqrt{\Xout^2 + (p\Ro)^2} \,.
      \label{eq_RinRout}
\end{equation}
The in- and outbound MC radius boundaries are shown in the bottom row of \fig{t2R} with $\By (R)$. The blank space around the magnetic cloud axis ($R=0$) is associated with the distance to the spacecraft at the closest approach. The last step is to define the inferior and superior values of $R$ at the MC boundaries:
   \begin{equation}
    \Rinf = \rm min(\Rin,\,\Rout) \rm \quad and \quad 
    \Rsup = max(\Rin,\,\Rout) \,.
      \label{eq_RinfRsup}
   \end{equation}

The examples shown in \fig{t2R} have $\Ro$ in between $\Rinf$ and $\Rsup$. This is not observed for all the cases \citep[see Fig. 2f of][]{Demoulin19}. In particular, 42\% of MCs in the set $\Qall$ have $\Rinf<\Rsup<\Ro$, so where $\Bz=0$ is located beyond the MC boundaries. At the opposite, 20\% have $\Rinf>\Rsup>\Ro$, so where the FR core ($\Bz>0$) is surrounded by $\Bz<0,$ so they are named  annulus cases \citep[see, e.g.][]{Vandas01}.
Finally, in \tab{catalogue} we present the results of $\sgn(\asf)$, $\Rinf$, $\Rsup$, and the values of $\chiR$ for the Lundquist fit with both possible signs of the $\asf$ for the set $\Qall$.

\begin{figure}[t!]            
\centering
\includegraphics[width=0.49\textwidth]{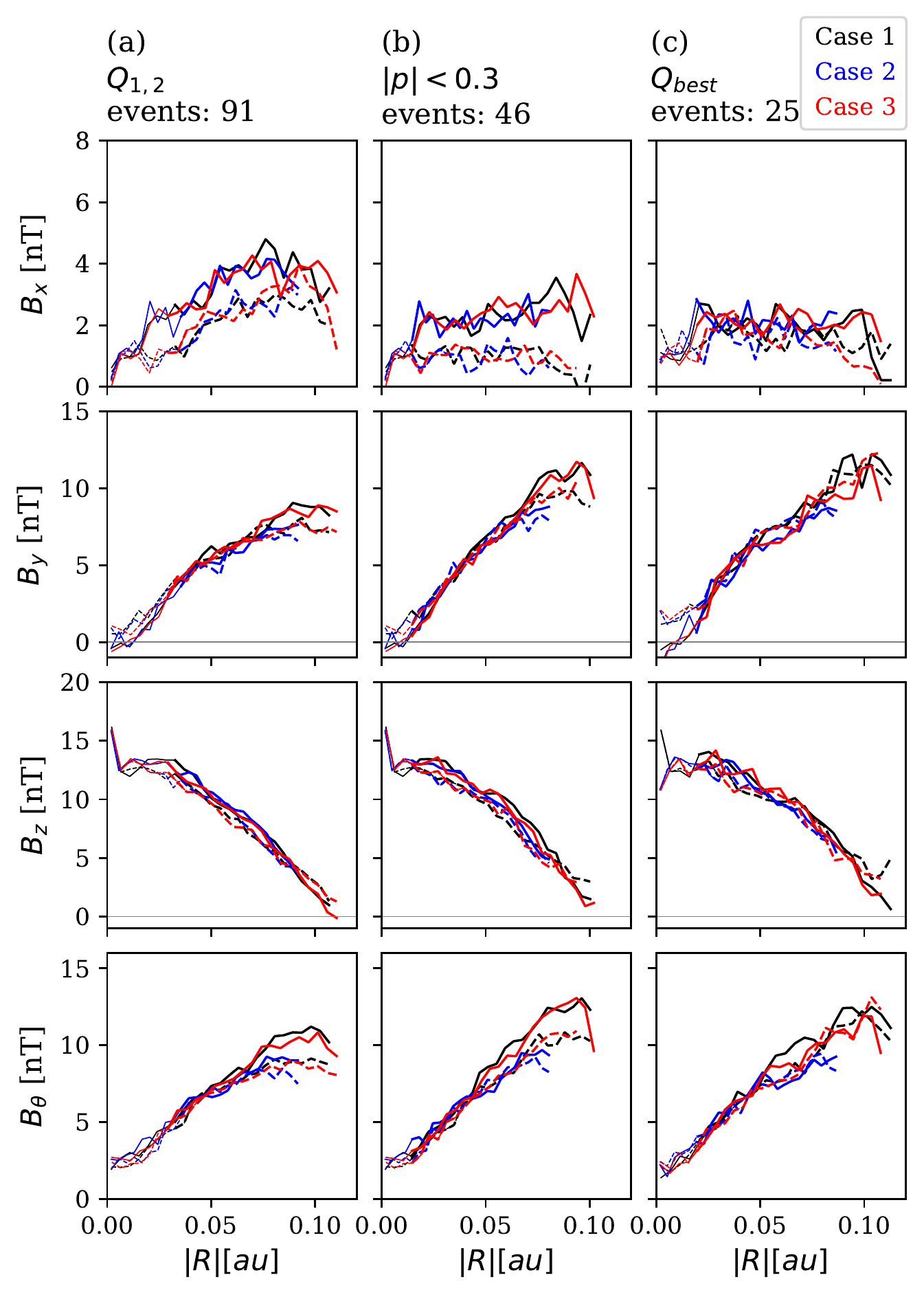}
\caption{Median profile of the magnetic field components derived with an SEA using first the MC data in $[-\Ro,\,\Ro]$ with $\Bz(\Ro)=0$ (black), second in the \insitu FR $[-\Rinf,\,\Rinf]$ (blue), and third in the non-eroded FR $[-\Rsup,\,\Rsup]$. The in- and outbound profiles are drawn with continuous and dashed lines, respectively.
The columns show three different samples: (a) the set $\Qall$;  (b) a subset of $\Qall$ with a low impact parameter ($|p|\leq 0.3$); and (c) a subset of $ \Qall$, $\Qbest$, with a symmetric magnetic field profile ($|\CB| \leq 0.05$) and a low impact parameter ($|p|\leq 0.3$).
}
 \label{fig_SEA2branch}
\end{figure}

\subsection{Superposed epoch analysis procedure}
\label{sect_SEA_method}

The SEA is a statistical tool useful to reinforce the common features between several individual MC profiles and minimise the peculiarities of each event. After defining the boundaries of the data to be superposed, an SEA of scalar quantities can be done. For vectors quantities, the data first need to be rotated in a common meaningful frame and defined with the same global sign so that the SEA of the components are meaningful.

In order to make a coherent superposition of the events, we rotated the magnetic field components into the FR frame defined so that $\Bz>0$ in the FR core. 
The sign of $\Bx$ and $\By$ in the FR frame depends on the sign of the FR helicity ($H$), so we first need to change all the MCs to positive helicity.  Next, the $\Bx$ sign depends on the sign of the closest approach distance ($CA$).  In summary, we change $\Bx$ to $\sgn(CA~H)~\Bx$ and $\By$ to $\sgn(H)~\By$ to have a set of FRs with positive helicity and positive closest approach distance.

Next, we need to define which is the relevant portion of the data to include for each MC. We investigated three possibilities that below we call the model, the \insitu, and the initial FR. \\
{\bf Case 1:} The first one used the Lundquist fitted model to define the FR boundaries where  $\Bz(\Ro)=0$. This can be partly justified by the presence in the corona of a potential field arcade above the erupting flux rope, which provides $\Bz =0$ after reconnection behind the FR \citep[this argument has its own limitations, see][]{Demoulin13}. This is the way Lepping built his MC tables and related papers. \\ 
{\bf Case 2:}   The second choice, or \insitu FR extension, was to consider only the FR crossed by the spacecraft, so as to analyse the data within $[-\Rinf, \Rinf]$. Two extreme interpretations of this choice are that this FR corresponds either to the remaining solar FR due to erosion, or to the restrictive region where an FR could be formed in the corona. In the latter case, the extra flux present on one side would be flux that could not be incorporated, via reconnection, in the FR (e.g. an overlying coronal magnetic arcade).  \\
{\bf Case 3:} Finally, a third choice, or non-eroded FR extension, was to consider all the MC data, so an FR defined within $[-\Rsup, \Rsup]$, considering that the flux missing on one side was lost by erosion during the travel. This missing flux is treated as data gap. Since the missing flux is present about in half MCs, either in the inbound region or outbound region \citep{Ruffenach15}, the corresponding SEA has approximately half the MCs near the interval boundaries.  \\
We also remind the reader that the central part of an MC, $[-p \Ro, p \Ro]$, has no data since the spacecraft does not explore the FR core. This implies that for all above boundary selection, the central part of the SEA is only realised with a few MCs (those observed with a low impact parameter $p$).

In order to superpose the FRs data on an SEA, the extension of the FRs should be the same. This is achieved for all MC data by normalising $R$ with $\Ro$, $\Rinf$, or $\Rsup$, respectively, for the three above-selected FR extensions. The result is an abscissa inside $[-1, 1],$ which allows us to superpose the various MC data.  

The computation of the SEA requires that each individual profile has the same number of data points. The number of data points increases with the duration of the observed MC. In order to obtain the same number of data points in every MC profile, we defined a grid of equally spaced bins, so that there were 25 bins within the inbound region and 25 bins within the outbound region of MCs. All data points that are present in each bin were averaged to a single value. 
Next, the average (mean or the median values) of the MC set was performed in each bin. This defined the SEA profile for each magnetic field component.  Here we used the median as it is more robust to outlier values \citep[see][for a comparison between mean and median results]{Regnault20,Lanabere20}.

The results of the above three SEAs cannot be directly compared since a different normalisation of $R$ was needed. 
We corrected this by multiplying the normalised radius by $\Romean$, $\Rinfmean$ and $\Rsupmean$, where $\mean$ is an average over the MCs considered. For $\Qall$ set $\Romean=0.109~\au$, $\Rinfmean = 0.094~\au$, and $\Rsupmean = 0.113~\au$, for $p<0.3$ set $\Romean=0.102~\au$, $\Rinfmean = 0.082~\au$, and $\Rsupmean = 0.104~\au$, while for $\Qbest$ $\Romean=0.115~\au$, $\Rinfmean = 0.088~\au$, and $\Rsupmean = 0.110~\au$. Then, the renormalisation of the SEA results to comparable sizes was a modest factor. First, the above normalisation of $R$ by $\Ro$, $\Rinf$, or $\Rsup$, and second the normalised radius was multiplied by $\Romean$, $\Rinfmean$, or $\Rsupmean$  to make the SEAs comparable and to provide a radius $R$ in $\au$ used in \fig{SEA2branch}. 

\subsection{Definition of MC subsets} 
\label{sect_Data_MC_selection}

The full set of MCs corresponds to 91 MCs from Lepping's catalogue with quality $\Qo=1$ and $\Qo=2$, defined as $\Qall$ in \sect{Data_MC}. Next, we defined new subsets according to the impact parameter $p$ and the asymmetry $\CB$.

The impact parameter $p$ expresses how far from the MC axis the spacecraft scans the MC. This implies that the azimuthal magnetic field components ($\Bx,\,\By$) are significantly dependant on the impact parameter $p$. In order to superpose the magnetic field components, with low $\Bx$ values and a better FR orientation, we selected MCs with $|p|\leq 0.3$. This kept almost half of the $\Qall$ events. Finally, the set defined by $\Qbest$ (\sect{asf_sign}) kept 25 events. It corresponds approximately to symmetric events with a low impact parameter, $|\CB| \leq 0.05$ and $|p| \leq 0.3$.

\subsection{Comparing SEA profiles} 
\label{sect_Comparing_SEA_profiles}

A comparison of the SEA profiles for the three sets of MCs ($\Qall$, $\Qall$ with $|p|\leq 0.3$, and $\Qbest$) is shown for three different FR boundaries: the model FR in [$-\Ro,\,\Ro$], the \insitu FR in [$-\Rinf,\,\Rinf$], and the non-eroded FR [$-\Rsup,\,\Rsup$].
In order to compare the asymmetry between the in- and outbound boundaries of the FR, we set the abscissa to $|R|$ and we flipped the $\By$ sign in the inbound boundary (to better compare it to the outbound values). 

The SEA of $\Bx$ is shown in the top row of \fig{SEA2branch}. With $\Qall$ set, $\Bx$ increases with $R$ up to $R\approx 0.07$. This contrasts with the flat profile of $\Bx$ obtained with $|p|\leq 0.3$ and $\Qbest$. In fact, the increasing profile with $\Qall$ set is due to a bias: only MCs observed with a low $p$ contribute near the origin, and a small $|p|$ implies small $\Bx$ values \eqp{bL_cartesian}. Progressively, for larger $|R|$, more MCs contribute with larger $|p|$, implying a larger $\Bx$ contribution.
Selecting MCs with $|p|\leq 0.3$ limits this bias to $|R|$ values closer to the origin (middle and right columns). Moreover, the slight decrease in $\Bx$ near the boundaries for FRs defined with $\Ro$ and $\Rsup$ could be due to perturbations by the surrounding medium present at the FR periphery, or the inclusion of non-MC data. We conclude that the results show a flat $\Bx$ profile when the biases are minimised. 

The profile of $\By$ is also affected by the $p$ bias, although differently than $\Bx$. Going more to the periphery (larger $|R|$), more MCs with larger $|p|$ contribute, so $\By$ is weaker for $\Qall$ compared with $|p| \leq 0.3$ and $\Qbest$ results. For all sets, $\By(0) \approx 0$ as expected in the FR model. The azimuthal component $\Btheta$ is computed from $\Bx$ and $\By$ assuming a circular FR cross section. Thus, it includes the biases of $\Bx$ and $\By$. For $|p| \leq 0.3$ and $\Qbest$, $\Btheta$ is almost a linear function of $R$. 

$\Bx$ is significantly stronger in the inbound region (continuous lines in \fig{SEA2branch}) than in the outbound region for the $\Qall$ and $|p|\leq 0.3$ sets. This is also true for $\By$ to a lesser extent for large $R$ values. This is due to the frequently stronger $B$ measured in the inbound region compared to the outbound one ($\CB$ is strongly negative for a fraction of MCs, \fig{asf}b). This effect disappears if we limit the $B$ asymmetry of the included MCs, as in the $\Qbest$ set. 

The profile of $\Bz$ is expected to be much less affected by the $|p|$ value, since $\Bz$ is maximum around the origin with a weak dependence on $R$ in the core. Indeed, $\Bz$ profiles in the third row of \fig{SEA2branch} are closer than for the other components.
Still, near $R=0$, there are too few MCs for the SEA to be reliable. Thus, the local $\Bz$ peak near the origin is due to a larger $B$ for the few MCs with low $|p|$ cases.  At the FR boundary, $\Bz$ is almost zero, especially for the $\Qall$ set.
 
Setting the FR boundary at $\Ro$, $\Rinf$, or $\Rsup$ before making the SEA finally has little effect on the results since the curves of these cases nearly superpose in \fig{SEA2branch}. Indeed, in the common core region, the same \insitu data are included and they are phased the same way (using the closest approach distance to separate in- and outbound regions). Thus, the difference between the three SEAs is related to how the binning is done (same number of SEA bins while the radial extension taken is different). As a result, the closeness of the curves in \fig{SEA2branch} is a logical consequence of the weak effect of the data rebinning. The main differences  between the results obtained with different selected FR boundaries are the large $|R|$ values, as expected.

The bias introduced by $p$ is partly solved by selecting MCs with low $|p|$ values with the limit of having enough MCs, at least a few tens, to perform an SEA that is not too affected by fluctuations (due to individual MC peculiarities). With these limitations, the $\Bx$ profile is almost flat, and both the $\By$ and $\Btheta$ profiles are close to being linear with the radius. This implies an axial current density that is uniform, such as in the model of \citet{Hidalgo02}.

In summary, when the effect of the various bias are well decreased, a typical profile of $B$ is obtained independently of the precise method used to derive it (right column of \fig{SEA2branch}). The selection of FR extension at $\Ro$, $\Rinf$, or $\Rsup$ does not significantly change the SEA profiles of the magnetic field components in the common region of the radius $R$. Using the FR extension up to $\Rsup$ extends the profiles to larger radii, with more fluctuations, since only about half of the MCs contribute to the largest $R$ values. It is much more significant to select a low impact parameter $|p|$ and an asymmetry $|\CB|$ to limit the corresponding biases.

\begin{figure}     
    \centering
    \includegraphics[width=0.49\textwidth]{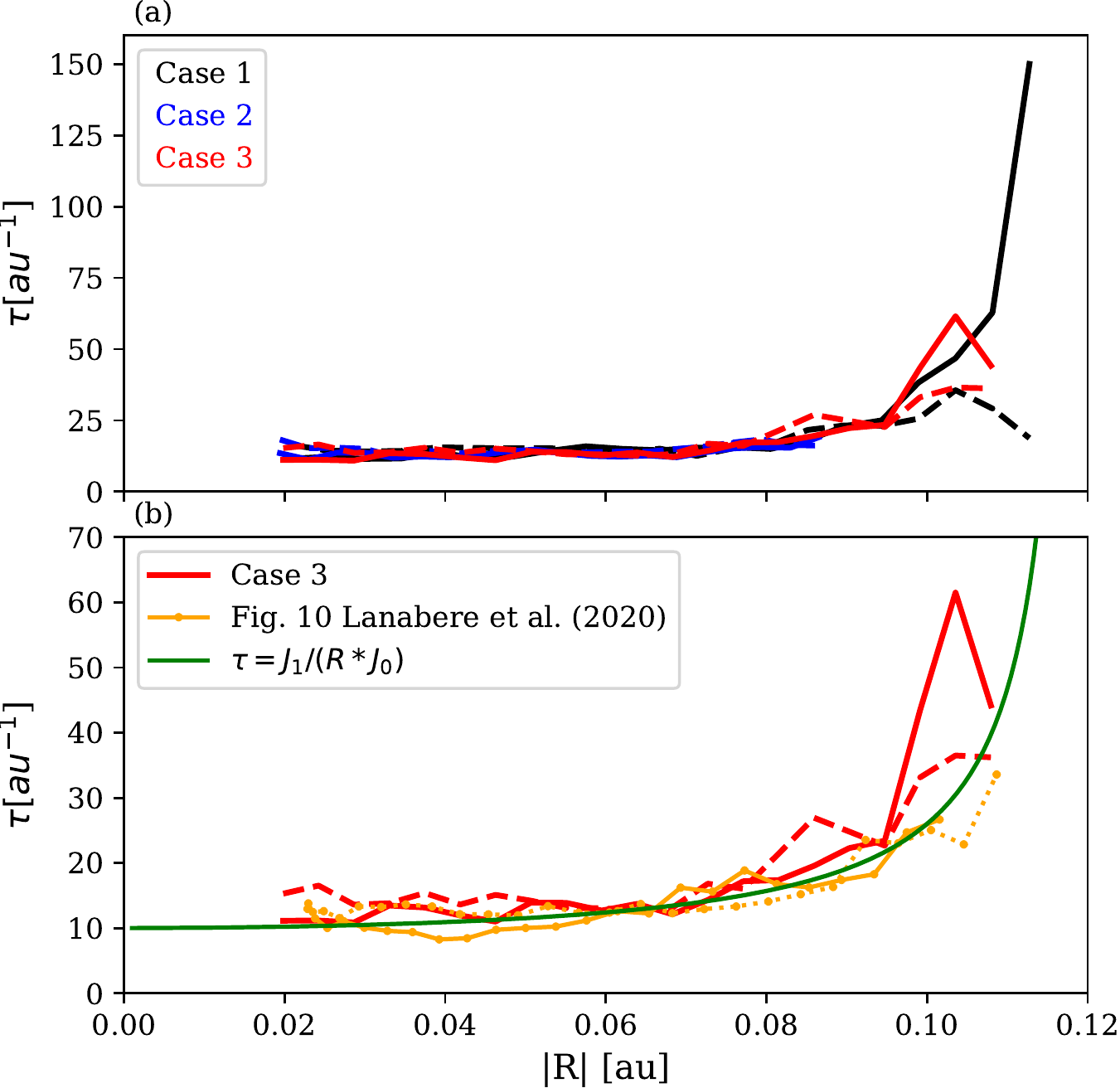}
    \caption{Twist profiles for $\Qbest$ set.
    (a) Comparison of the in- and outbound profiles for the three cases: the Lepping FR extension $[-\Ro,\Ro]$ (black, case 1), the \insitu FR extension $[-\Rinf,\Rinf]$ representing the eroded FR observed at 1 au (blue, case 2), and the non-eroded FR extension $[-\Rsup,\Rsup]$ representing the possible initial FR (case 3). The in- and outbound profiles are represented with solid and dashed lines, respectively. 
    (b) Comparison of case 3 ($\Rsup$ boundary) with the results of Fig. 10 in \citet{Lanabere20}.  Their SEA results for in- and outbound (in orange) and the Lundquist model with $\Ro =0.12$ au (in green) are included.}
    
    \label{fig_twist_comparison}
\end{figure}

\section{Generic twist profile}
\label{sect_SEA_twist}

A main characteristic of an FR is its amount of magnetic twist, which is defined as the amount of rotation that the magnetic field lines make around the FR axis per unit length along the axis. The twist distribution can be computed considering a cylindrical symmetry as $\tau = B_\theta/(R\Bz)$ in radian/au (or simply au$^{-1}$), with R the radial distance from the FR axis.

The twist as a function of $R$ for the three subsets of \sect{Data_MC_selection} were computed following the procedure described in \citet{Lanabere20}. In order to show a representative twist profile for each subset, the twist is shown where more than $50\%$ of the events contribute to the SEA (so that there is a limited variation in the statistical weight across the twist profile).

The twist profiles derived with the SEA extension set to the model (case 1), \insitu (case 2), and non-eroded FR (case 3) are compared in \fig{twist_comparison}a for $\Qbest$ set (to avoid biases as much as possible).
The twist profile is almost constant in the FR core ($|R|<0.06~\au$) and comparable within the in- and outbound profiles.
Indeed, the average of the in- and outbound profiles are $\tau=13.85, 13.82,$ and $13.4 ~\au^{-1}$ for cases 1 to 3, respectively.
This slight difference is the result of the sensitivity of the twist profile to the SEA binning, to magnetic fluctuations in MCs, and to the low number of MCs included, at most 25 events, to avoid biases (\sect{Comparing_SEA_profiles}).

The twist profile near the origin could not be reliably determined as there were too few MCs observed with a small impact parameter. Additionally, the twist determination implies the ratio of two small quantities ($B_\theta$ and $R$) in the vicinity of the FR axis. Since $R$ is vanishing by definition of the FR axis, while $B_\theta$ incorporates the fluctuations of $\Bx$ and $\By$, $B_\theta$ is typically not vanishing in the data of individual MCs, not even in SEAs (last row of \fig{SEA2branch}). This implies that the computed twist profile typically diverges near the FR axis.  In fact, the twist profile cannot be reliably derived there with an SEA, and even less so for an individual MC. 

For $|R|>0.06~\au$, the twist has a small increase with $|R|$, which is coherent between the cases up to $|R| \approx 0.09~\au$. For larger $|R|$ values, in cases 1 and 3 there is a large dispersion of the results.  This dispersion is enhanced by the low $\Bz$ value reached in the inbound profile of case 1 (\fig{SEA2branch}, right column). On the other side, $\Bz$ has a local increase for case 1 at the outbound profile, which implies a decrease in the twist there. The results of case 3 is in between these two extremes (\fig{twist_comparison}a). Selecting the set $\Qall$ or $|p|\leq 0.3$ would lead to different values of $\Bz$ near the FR boundary. This implies important variations in the deduced twist. We conclude that the twist significantly increases at the FR periphery, but its value has a large uncertainty. 

We next compared the present results to the ones of \citet{Lanabere20} in \fig{twist_comparison}b. The global twist trend is comparable.   This justifies a posteriori the results of \citet{Lanabere20} where the SEA was simply performed using the central time, and not the time of the minimum distance approach to the FR axis, as in this study, so finally using $\Rin$ and $\Rout$ even if they differ. The similarity of the results is at first surprising because SEA could accumulate the bias of a dominant erosion from one side.  In that case, there would be a systematic shift of the MC and the FR boundaries, and so different SEA profiles. In fact, MCs are statistically eroded almost symmetrically at the front and rear \citep{Ruffenach15}, as they are present in approximately equal amounts in our various samples, so that a systematic bias of one side erosion has a weak signature in the SEA using MC boundaries. This is the strength of the SEA to erase the peculiarities (here erosion side and amount) of individual cases. Finally, the above uniform twist values are larger than previous studies, with $\tau \sim 9.4$ au$^{-1}$ \citep[][performed with a Grad-Shafranov model applied to an MC crossed by two spacecraft]{Mostl09}, and with $\tau \sim 11.5~\au^{-1}$ \citep[][derived with an SEA analysis]{Lanabere20} .

Finally, we compared the results to the Lundquist model (green curve in \fig{twist_comparison}b).  
The SEA twist is even flatter than the Lundquist's twist in the FR core, especially for cases 1-3.  Within the SEA twist dispersion discussed above, the SEAs show a comparable increase in the twist at the FR periphery.
Thus, we confirm that the Lundquist's model well represents the median twist of MCs.

\section{Summary and conclusions}
\label{sect_Conclusions}
    
Previous studies have shown contrasted results for the magnetic twist profiles present in MCs, even in the framework of a supposed locally cylindrical symmetric magnetic field (\sect{Introduction}).  The twist profile is especially delicate to derive both because the data need to be precisely rotated in the FR frame and because the twist is derived from the ratio of two quantities that vanish at the FR axis.
A way to improve the derived twist profile is to use the powerful SEA method, which weakens the individual MC peculiarities and decreases fluctuations.  
This was done in a simple way by \citet{Lanabere20}, choosing the central time of the time series to split the in- and outbound branches, and also using the MC boundaries to superpose the magnetic field in the FR frame.
However, since MCs are differently eroded during their travel from the Sun, one may wonder how justified it is to superpose data where the FR magnetic flux was partly removed on either side.   Thus, one of the main aims of this paper was to correct this possible bias.    

Another aim of the present study was to search, then to correct, other possible biases present within the SEA.    For that, we limited our study to the MCs of best quality present in Lepping's list.
The fitted Lundquist model provides the FR frame where the data are rotated in order to superpose, in a consistent way, the magnetic field components (azimuthal and axial components). We remind the reader that comparable results have been obtained with FR orientation defined with the minimum variance method \citep{Lanabere20}.
The magnetic components also need to add up. This requires that all cases have the same sign of magnetic helicity and impact parameter.  Furthermore, the magnetic signals need to be phased in time so that the closest approach distance of all the FRs 
are synchronised to a common reference time.
This is typically not the case when the MC boundaries are used for the SEA \citep{Lanabere20} since MCs are typically eroded on one or both sides when observed at 1 au. The Lepping's list does not provide the time of the axis closest approach because it provides only the absolute value of a related parameter, called the $\asf$ \citep[see,][]{Demoulin19}.  
We thus performed a least square fit, as originally done, to recover the sign of $\asf$.
This is a useful byproduct obtained in the present study, and presented in \tab{catalogue}. 
This fully completed the parameter set of the best Lundquist model associated with each MC.  

In order to have a more coherent superposition of the magnetic field components in the SEA, we used the fitted model to align the various parameter data. We selected three possible FR boundaries (see \fig{schema}).
The first one was set at $\Ro$, where $\Bz(\Ro)=0$ for the Lundquist model (original Lepping's choice).  The second one, $\Rinf$, was set to include only the inner spatially symmetric part of the FR extension, crossed by the spacecraft, so it corresponds to the FR remaining when crossed by the spacecraft. Extra magnetic flux present before or after was interpreted as the remaining magnetic flux on the side opposite to erosion (rather than extra flux not belonging to the FR close to the Sun). The third radius, $\Rsup$ was designed to also incorporate this extra magnetic field at the outer part. 
This implies a data gap on the outer part of the other side. Since erosion occurred almost as frequently at the front and the rear \citep{Ruffenach15}, the periphery part of the profiles were based on about half MC data.

Even selecting the best observed MCs, of quality $\Qall$, within Lepping's list, biases were present in the corresponding SEA.
A first bias was tied to the magnitude of the impact parameter $p$. The largest effect was an increasing profile of $\Bx$ with distance to the FR axis. This bias was strongly weakened when the impact parameter was limited to $|p|\leq 0.3$ (\fig{SEA2branch}). 
A second bias was linked to the field strength asymmetry between the in- and outbound regions.  This bias can affect the quality of the obtained FR orientation, with consequences on possible mixing of B components in the local FR frame.  This asymmetric effect is again mostly present in the $\Bx$ component, and could be reduced by limiting the B asymmetry of the MCs included within the SEA. 
Finally, a third possible bias was linked to flux  erosion during the travel from the Sun. This bias was found to be weak as the profile of the B components were similar with the data limited to the MC or by a given radius of the fitted FR. 
All these results justify and set on firmer ground the results of \cite{Lanabere20} obtained by directly including the MC data in a simple way, neglecting the in-out bound asymmetries present in each of the analysed events.

The twist profile within an MC is typically difficult to derive because the spacecraft typically crosses only a part of the FR, in general missing the core, and because the computations need to be done in the FR frame with an axisymmetric hypothesis.  Magnetic fluctuations and peculiarities of individual MCs worsen the derivation of the twist profile. The expected singularity of the twist profile at the FR radius (with finite B fluctuations) and at its periphery (where the axial field component vanishes) worsen the derivation of the twist profile. These difficulties soften with an SEA. The various derived profiles, with a different selected outer FR radius and a different set of MCs (to analyse bias effects) point to a flat twist profile in the FR core (about two-thirds of the FR radius).
In the vicinity of the FR axis, there are presently not enough MCs observed with a low impact parameter to get a reliable twist profile.
At the FR periphery, the twist sharply increases, while the uncertainty is large mostly due to a weak axial B component easily affected by external perturbations.
This implies that the original FR present close to the Sun is expected to have a more strongly twisted periphery, in agreement with the solar FR further building up during eruptions \citep[see][]{Demoulin19}. 

To summarise, we verify that SEA of MCs is weakly sensible to the precise definition of the boundaries considered to do the superposition. Still, SEA includes the biases present in the original data set. 
The selection of the included MCs can significantly reduce these biases, but at the cost of decreasing the number of events and consequently having weaker statistics, thus including more fluctuations in the SEA profiles.
Further improvements would require a larger data base of observed MCs, especially to better constrain the twist profile near the centre and the periphery of the FR).

\begin{acknowledgements}
S.D. and V.L. acknowledge partial support from the Argentinian grants PICT-2019-02754 (FONCyT-ANPCyT) and UBACyT-20020190100247BA (UBA).
P.D. thank the Programme National Soleil Terre of the CNRS/INSU for financial support.
S.D. is member of the Carrera del Investigador Cien\-t\'\i fi\-co, CONICET.
We recognise the collaborative and open nature of knowledge creation and dissemination, under the control of the academic community as expressed by Camille No\^us at http://
www.cogitamus.fr/indexen.html.
\end{acknowledgements}

%
\bibliographystyle{aa}
\bibliography{mc}

\clearpage
\onecolumn

\begin{appendix}
\section{Table of analysed MCs, including the obtained $\sgn(\asf)$.}
\label{sect_Table}

\begin{longtable}{rrrccccccc}
\caption{Results for Lepping MCs events of quality $\Qall$.}\\
\hline\hline
\label{tab_catalogue}
Event & Code & $|\asf$| & $\sgn(\asf)$& $\Rinf/\Ro$ & $\Rsup/\Ro$& $\Ro$& $\chi_{R,minus}$ & $\chi_{R,plus}$ & $\Qbest$\\
\hline
\endfirsthead
\caption{continued.}\\
\hline\hline
Event & Code & $|\asf|$ & $\sgn(\asf)$& $\Rinf/\Ro$ & $\Rsup/\Ro$& $\Ro$& $\chi_{R,minus}$ & $\chi_{R,plus}$ & $\Qbest$\\
\hline
\endhead
\hline
\endfoot
\hline
\endlastfoot
1       &       1       &       17.4&   +       &       0.860   &       1.118   &       0.108   &       0.43    &       0.27    &       0       \\
2       &       2.2     &       6.8     &       +       &       0.929   &       0.987   &       0.152   &       0.31    &       0.27    &       0       \\
3       &       3       &       1.0     &       +       &       1.208   &       1.222   &       0.042   &       0.15    &       0.15    &       0       \\
4       &       5       &       26.5&   +       &       0.734   &       1.070   &       0.126   &       0.50    &       0.28    &       0       \\
5       &       6       &       2.3     &       +       &       1.050   &       1.099   &       0.129   &       0.16    &       0.15    &       1       \\
6       &       8       &       27.5&   +       &       0.679   &       1.184   &       0.175   &       0.52    &       0.27    &       0       \\
7       &       9       &       6.8     &       +       &       0.806   &       0.920   &       0.086   &       0.28    &       0.28    &       1       \\
8       &       10      &       10.3&   -       &       0.898   &       1.053   &       0.107   &       0.27    &       0.30    &       0       \\
9       &       11      &       4.2     &       +       &       1.036   &       1.108   &       0.143   &       0.19    &       0.17    &       0       \\
10      &       12      &       14.3&   +       &       0.921   &       1.225   &       0.095   &       0.31    &       0.13    &       1       \\
11      &       14.1&   6.3     &       +       &       0.636   &       0.689   &       0.131   &       0.36    &       0.34    &       0       \\
12      &       15      &       19.3&   +       &       0.754   &       1.084   &       0.093   &       0.37    &       0.20    &       0       \\
13      &       17      &       5.2     &       -       &       0.814   &       0.866   &       0.093   &       0.25    &       0.28    &       0       \\
14      &       22      &       7.4     &       +       &       0.601   &       0.697   &       0.117   &       0.11    &       0.08    &       1       \\
15      &       23      &       12.0&   -       &       0.728   &       0.859   &       0.198   &       0.30    &       0.38    &       0       \\
16      &       24      &       4.1     &       +       &       1.152   &       1.227   &       0.114   &       0.15    &       0.14    &       0       \\
17      &       25      &       1.8     &       -       &       0.577   &       0.597   &       0.110   &       0.11    &       0.11    &       1       \\
18      &       26      &       1.6     &       -       &       0.899   &       0.920   &       0.058   &       0.24    &       0.25    &       0       \\
19      &       28      &       1.5     &       +       &       0.928   &       0.957   &       0.119   &       0.15    &       0.14    &       1       \\
20      &       30      &       5.5     &       -       &       1.361   &       1.454   &       0.147   &       0.12    &       0.18    &       0       \\
21      &       31      &       5.5     &       +       &       0.880   &       0.982   &       0.164   &       0.18    &       0.16    &       1       \\
22      &       33      &       22.7&   +       &       0.589   &       0.902   &       0.035   &       0.35    &       0.19    &       0       \\
23      &       34      &       24.3&   -       &       0.716   &       1.136   &       0.104   &       0.13    &       0.39    &       1       \\
24      &       35      &       4.4     &       +       &       1.081   &       1.179   &       0.107   &       0.31    &       0.28    &       0       \\
25      &       36      &       14.8&   +       &       0.965   &       1.194   &       0.204   &       0.27    &       0.14    &       0       \\
26      &       38      &       5.3     &       -       &       1.097   &       1.217   &       0.124   &       0.18    &       0.21    &       0       \\
27      &       41      &       9.5     &       +       &       0.873   &       1.041   &       0.136   &       0.28    &       0.21    &       1       \\
28      &       44.3&   20.8&   +       &       0.719   &       0.954   &       0.132   &       0.34    &       0.19    &       0       \\
29      &       47      &       33.0&   -       &       0.683   &       1.005   &       0.115   &       0.28    &       0.29    &       0       \\
30      &       49      &       14.2&   -       &       0.945   &       1.257   &       0.142   &       0.21    &       0.36    &       0       \\
31      &       51      &       1.1     &       -       &       1.021   &       1.043   &       0.092   &       0.16    &       0.16    &       1       \\
32      &       52      &       16.7&   +       &       0.572   &       0.794   &       0.119   &       0.24    &       0.15    &       1       \\
33      &       54      &       22.4&   -       &       0.658   &       1.011   &       0.138   &       0.13    &       0.34    &       1       \\
34      &       55.1&   0.8     &       -       &       0.851   &       0.864   &       0.083   &       0.14    &       0.15    &       1       \\
35      &       56      &       21.1&   -       &       0.956   &       1.072   &       0.191   &       0.17    &       0.19    &       0       \\
36      &       57      &       9.2     &       +       &       0.774   &       0.813   &       0.125   &       0.14    &       0.10    &       0       \\
37      &       58      &       28.9&   +       &       0.625   &       1.130   &       0.133   &       0.43    &       0.12    &       1       \\
38      &       59      &       15.3&   +       &       0.685   &       0.859   &       0.116   &       0.21    &       0.11    &       0       \\
39      &       60      &       22.5&   +       &       0.713   &       1.031   &       0.126   &       0.29    &       0.18    &       0       \\
40      &       61      &       16.4&   -       &       1.182   &       1.570   &       0.127   &       0.29    &       0.45    &       0       \\
41      &       64      &       17.7&   -       &       0.472   &       0.650   &       0.104   &       0.09    &       0.17    &       1       \\
42      &       65      &       3.2     &       +       &       0.993   &       1.058   &       0.216   &       0.23    &       0.22    &       1       \\
43      &       66      &       1.4     &       -       &       0.795   &       0.807   &       0.159   &       0.19    &       0.20    &       0       \\
44      &       68      &       5.2     &       +       &       1.058   &       1.082   &       0.212   &       0.09    &       0.09    &       0       \\
45      &       71      &       8.4     &       -       &       0.544   &       0.640   &       0.127   &       0.09    &       0.12    &       1       \\
46      &       72.1&   24.8&   +       &       0.696   &       1.052   &       0.073   &       0.34    &       0.27    &       0       \\
47      &       73      &       1.6     &       +       &       1.015   &       1.034   &       0.105   &       0.17    &       0.17    &       0       \\
48      &       76      &       27.0&   +       &       0.497   &       0.847   &       0.144   &       0.37    &       0.24    &       0       \\
49      &       77      &       14.4&   +       &       1.049   &       1.401   &       0.090   &       0.43    &       0.23    &       0       \\
50      &       78      &       2.6     &       -       &       1.045   &       1.089   &       0.197   &       0.26    &       0.28    &       0       \\
51      &       80      &       21.5&   -       &       0.804   &       1.192   &       0.178   &       0.16    &       0.41    &       1       \\
52      &       81      &       5.4     &       -       &       0.793   &       0.884   &       0.120   &       0.15    &       0.19    &       1       \\
53      &       82      &       1.0     &       -       &       0.939   &       0.956   &       0.087   &       0.12    &       0.12    &       0       \\
54      &       83      &       6.7     &       -       &       0.862   &       0.964   &       0.072   &       0.21    &       0.24    &       0       \\
55      &       84      &       1.9     &       +       &       0.859   &       0.885   &       0.075   &       0.12    &       0.11    &       0       \\
56      &       85      &       0.5     &       +       &       1.140   &       1.147   &       0.195   &       0.13    &       0.13    &       0       \\
57      &       86      &       13.1&   +       &       1.134   &       1.449   &       0.090   &       0.38    &       0.30    &       0       \\
58      &       87      &       28.8&   +       &       0.520   &       0.840   &       0.131   &       0.27    &       0.12    &       0       \\
59      &       89      &       9.9     &       +       &       0.892   &       1.043   &       0.074   &       0.23    &       0.19    &       0       \\
60      &       92      &       18.7&   +       &       0.883   &       1.289   &       0.068   &       0.42    &       0.22    &       0       \\
61      &       94      &       18.1&   +       &       0.640   &       0.891   &       0.113   &       0.21    &       0.14    &       1       \\
62      &       95      &       2.6     &       -       &       1.004   &       1.044   &       0.067   &       0.17    &       0.17    &       0       \\
63      &       99      &       4.5     &       +       &       0.928   &       1.011   &       0.046   &       0.20    &       0.20    &       1       \\
64      &       100     &       3.0     &       -       &       1.131   &       1.200   &       0.066   &       0.30    &       0.32    &       0       \\
65      &       102     &       23.0&   -       &       0.552   &       0.826   &       0.066   &       0.10    &       0.27    &       1       \\
66      &       104     &       14.0&   +       &       0.659   &       0.872   &       0.060   &       0.28    &       0.26    &       0       \\
67      &       105     &       9.0     &       -       &       1.124   &       1.335   &       0.085   &       0.23    &       0.32    &       0       \\
68      &       107     &       2.9     &       +       &       0.845   &       0.863   &       0.091   &       0.32    &       0.30    &       0       \\
69      &       108     &       34.5&   -       &       0.414   &       0.850   &       0.034   &       0.15    &       0.42    &       1       \\
70      &       109     &       23.4&   -       &       0.616   &       0.855   &       0.061   &       0.13    &       0.27    &       0       \\
71      &       110     &       16.5&   -       &       0.521   &       0.719   &       0.033   &       0.17    &       0.27    &       0       \\
72      &       113     &       6.8     &       +       &       0.862   &       0.900   &       0.168   &       0.36    &       0.34    &       0       \\
73      &       119     &       15.2&   +       &       0.937   &       1.266   &       0.092   &       0.34    &       0.19    &       0       \\
74      &       120     &       0.0     &       +       &       1.102   &       1.102   &       0.144   &       0.54    &       0.54    &       0       \\
75      &       123     &       2.1     &       +       &       0.878   &       0.893   &       0.055   &       0.22    &       0.22    &       0       \\
76      &       131     &       23.3&   +       &       0.664   &       1.053   &       0.174   &       0.33    &       0.14    &       1       \\
77      &       132     &       10.7&   -       &       0.992   &       1.229   &       0.057   &       0.20    &       0.29    &       0       \\
78      &       133     &       30.6&   +       &       0.555   &       1.033   &       0.046   &       0.42    &       0.23    &       0       \\
79      &       136     &       24.4&   -       &       0.641   &       1.014   &       0.106   &       0.30    &       0.48    &       0       \\
80      &       137     &       17.3&   +       &       1.218   &       1.680   &       0.017   &       0.53    &       0.22    &       0       \\
81      &       142     &       0.7     &       -       &       0.988   &       1.001   &       0.080   &       0.21    &       0.22    &       1       \\
82      &       144     &       4.8     &       +       &       1.001   &       1.099   &       0.111   &       0.24    &       0.21    &       0       \\
83      &       145     &       4.0     &       -       &       0.831   &       0.877   &       0.078   &       0.46    &       0.49    &       0       \\
84      &       146     &       20.0&   -       &       0.993   &       1.056   &       0.101   &       0.18    &       0.26    &       0       \\
85      &       148     &       5.7     &       +       &       0.898   &       1.005   &       0.069   &       0.20    &       0.17    &       0       \\
86      &       150     &       4.4     &       -       &       0.886   &       0.951   &       0.070   &       0.17    &       0.18    &       0       \\
87      &       151     &       21.3&   +       &       1.000   &       1.066   &       0.100   &       0.58    &       0.50    &       0       \\
88      &       153     &       36.6&   -       &       0.566   &       0.883   &       0.057   &       0.33    &       0.53    &       0       \\
89      &       158     &       9.3     &       +       &       0.895   &       1.054   &       0.125   &       0.31    &       0.25    &       0       \\
90      &       159     &       27.5&   -       &       0.795   &       1.244   &       0.094   &       0.16    &       0.47    &       0       \\
91      &       160     &       9.2     &       +       &       1.142   &       1.370   &       0.093   &       0.30    &       0.22    &       0
\end{longtable}
\tablefoot{The event Lepping's code, $|\asf|$, $\Ro$, and $\Qbest$ are from Lepping's table.  The sign of the $\asf$ was obtained by selecting the case with minimum $\chi_{R}$. $\Rinf$ and $\Rsup$ are defined in \eq{RinfRsup}.}

\end{appendix}

\end{document}

%% file: definitions.tex
\definecolor{darkgreen}{rgb}{0.0, 0.5, 0.0}
\definecolor{orange}{rgb}{1.0, 0.4,0.0}

\newcommand{\BE}{\begin{equation}}
\newcommand{\EE}{\end{equation}}
\newcommand{\BA}{\begin{eqnarray}}
\newcommand{\EA}{\end{eqnarray}}
 \newcommand{\fig}[1]{Fig.~\ref{fig_#1}}
 \newcommand{\figsss}[1]{Figure~\ref{fig_#1}}
 \newcommand{\figs}[2]{Figs.~\ref{fig_#1} and \ref{fig_#2}}

 \newcommand{\sect}[1]{Sect.~\ref{sect_#1}}

 \newcommand{\eq}[1]{Eq.~(\ref{eq_#1})}
  \newcommand{\eqp}[1]{(Eq.~\ref{eq_#1})}

\newcommand{\insitu}{{in situ }}

\newcommand{\tab}[1]{Table~\ref{tab_#1}}

\newcommand{\rmd}{{\rm d }}

\newcommand{\uvec}[1]{\hat{\bf #1}}

\newcommand{\asf}{{\rm asf}}

\newcommand{\Btheta}{B_{\rm \theta}}
\newcommand{\bthetaL}{b_{\rm \theta, L}}
\newcommand{\bxl}{b_{\rm x, L}}
\newcommand{\byl}{b_{\rm y, L}}
\newcommand{\bzl}{b_{\rm z, L}}
\newcommand{\Bo}{B_{0}}

\newcommand{\Bx}{B_{\rm x}}
\newcommand{\By}{B_{\rm y}}
\newcommand{\Bz}{B_{\rm z}}
\newcommand{\bx}{b_{\rm x}}
\newcommand{\by}{b_{\rm y}}
\newcommand{\bz}{b_{\rm z}}

\newcommand{\vBL}{\vec{B}_L}
\newcommand{\vbL}{\vec{b}_L}

\newcommand{\CB}{C_{\rm B}}  
\newcommand{\chiR}{\chi_{\rm R}}

\newcommand{\au}{\rm au}
\newcommand{\mean}{\langle \rangle}
\newcommand{\pL}{\phi_{\rm L}}
\newcommand{\Qo}{Q_{\rm 0}} 
\newcommand{\Qall}{Q_{\rm 1,2}} 
\newcommand{\Qbest}{Q_{\rm best}} 

\newcommand{\Ro}{R_{\rm 0}}
\newcommand{\Rin}{R_{\rm in}}
\newcommand{\Rout}{R_{\rm out}}
\newcommand{\Rinf}{R_{\rm inf}}
\newcommand{\Rsup}{R_{\rm sup}}
\newcommand{\Romean}{\langle R_{\rm 0} \rangle}
\newcommand{\Rinfmean}{\langle R_{\rm inf} \rangle}
\newcommand{\Rsupmean}{\langle R_{\rm sup} \rangle}

\newcommand{\sgn}{{\rm sgn}} 
\newcommand{\tp}{t_{0,\rm plus}} 
\newcommand{\tm}{t_{0,\rm minus}} 
\newcommand{\tL}{\theta_{\rm L}}

\newcommand{\Vmean}{\langle V \rangle}

\newcommand{\uyGSE}{\uvec{y}_{\rm GSE}}
\newcommand{\uzGSE}{\uvec{z}_{\rm GSE}}
\newcommand{\utheta}{\uvec{e}_{\theta}}
\newcommand{\uz}{\uvec{e}_z}
\newcommand{\ur}{\uvec{e}_r}
\newcommand{\ux}{\uvec{e}_x}
\newcommand{\uy}{\uvec{e}_y}

\newcommand{\Xin}{X_{\rm in}}
\newcommand{\Xout}{X_{\rm out}}

